\documentclass{article}
\usepackage{graphicx} % Required for inserting images

\usepackage[nonatbib,preprint]{tackling_climate_workshop_style}
\usepackage[square,numbers]{natbib}
\usepackage[utf8]{inputenc} % allow utf-8 input
\usepackage[T1]{fontenc}    % use 8-bit T1 fonts
\usepackage{hyperref}       % hyperlinks
\usepackage{url}            % simple URL typesetting
\usepackage{booktabs}       % professional-quality tables
\usepackage{amsfonts}       % blackboard math symbols
\usepackage{nicefrac}       % compact symbols for 1/2, etc.
\usepackage{microtype}      % microtypography

\usepackage{subcaption}
\usepackage{wrapfig}
\usepackage{appendix}

\title{Spatio-temporal precipitation downscaling at hourly resolution based on km-scale regional simulations}
\author{%
  Henry Addison \\
  University of Bristol, Bristol, UK \\
  \texttt{henry.addison@bristol.ac.uk} \\
  \And
  Elizabeth J. Kendon \\
  Met Office Hadley Centre, Exeter, UK \\
  University of Bristol, Bristol, UK \\
  \And
  Peter A. G. Watson \\
  University of Bristol, Bristol, UK \\
}

\begin{document}

\maketitle

%% Instructions %%
% Submissions are limited to four pages for the Papers track
% References do not count towards this total. 
% Supplementary appendices are allowed but will be read at the discretion of the reviewers.

\begin{abstract}
% The abstract must be limited to one paragraph.
Effectively preparing for changes in extreme flooding requires understanding present and future rainfall extremes with return periods up to \(\sim\)100-1000 years, including changes in intensity, spatial structure at km-scale and sub-daily temporal variability. Conventional km-scale climate models give greatly increased realism compared to typical global climate models (GCMs) but are too computationally costly to allow enough simulations for robustly quantifying changes in extremes.
This work presents an approach based on video diffusion to generate realistic precipitation at 8.8km-hourly resolution by spatially and temporally downscaling daily, GCM-scale atmospheric variables. The emulator is trained to produce similar output to a \(4\times\)-coarsening of the Met Office’s 2.2km-resolution UK convection-permitting model. An application for such an emulator is to downscale large-ensemble GCM simulations to allow quantification of \(\sim\)100–1000-year return level rainfall, which in turn can be used to drive hydrological and flood models for improved projections of future changes in flood risk.
\end{abstract}

\section{Introduction}
\label{sec:intro}

A major societal challenge is enhancing resilience to the impacts of extreme rainfall, such as flooding, with climate change causing increases in intense precipitation in many regions.
For example, flooding produced by extreme rainfall causes damage costing \(\sim\)£500 million/year in the UK and this is projected to increase in future \cite{sayers2020third, batesClimateconditionedCatastropheRisk2023}.
Flood defences are often built to withstand extreme flood events at or beyond the 1-in-100 year return level \cite{ea2014flood, ea2025waterresourcesframework}. 
However, existing predictions of UK flood risk are unreliable. 
Major reasons for this are reliance on rainfall simulations from low-resolution climate models and simplified statistical approaches to flood prediction \cite[e.g.][]{daleManagingEffectsExtreme2021}.
This issue is further compounded by the failure of these approaches to capture changes in risk in the future due to the changing behaviour of storms (e.g. increased flash flood risk from summer downpours becoming shorter and more intense  \cite{kendonHeavierSummerDownpours2014, kendonUpdateUKCPLocal2021}).
This can be corrected by using predictions of km-scale, sub-daily rainfall from physically-based models, which are derived from fundamental laws \cite{batesClimateconditionedCatastropheRisk2023, archerFutureChangeUrban2024}. 
However, the increased realism of such climate models comes with a very high computational cost, preventing their use to produce the large simulation datasets required to quantify extreme rainfall and flood risks.

To address this challenge, machine learning (ML) has been shown to be effective for emulating dynamical downscaling climate models on daily time scales \cite[e.g.][]{lopez-gomezDynamicalgenerativeDownscalingClimate2025,addisonMachineLearningEmulation2026}. Deep generative models (DGMs), in particular diffusion models \cite{sohl-dicksteinDeepUnsupervisedLearning2015,songScoreBasedGenerativeModeling2021, hoDenoisingDiffusionProbabilistic2020}, are important for the highly stochastic problem of downscaling precipitation \cite{rampal2026intercomparisongenerativemachinelearning, rampal2026cordex}.
Previous efforts have successfully used DGMs to generate km-scale, subdaily precipitation via spatial-only downscaling \cite[e.g.][]{harrisGenerativeDeepLearning2022, rampalReliableGenerativeAdversarial2025, Leinonen_2020}.
However, the subdaily outputs of coarser models to be eventually downscaled are either not available or lack realism \cite{kendonRealismRainfallVery2012}.
    Previous spatio-temporal downscaling works have also used video super-resolution techniques, which model temporal consistency between frames as well as producing high quality spatial detail conditioned on a coarse version of the target \cite[e.g.][]{glawionSpateGANSpatioTemporalDownscaling2023}.
However, km-scale models can make changes to rainfall as coarse as well as fine-scales.
This limits the effectiveness of super-resolution approaches to downscale the outputs of coarser models.
Furthermore, \citet{rampalExtrapolationGenerativeAdversarial2024} show emulators lack skill when extrapolating to future climates, a problem for emulators trained only on observational or reanalysis datasets.

% This problem requires temporal as well as spatial downscaling and so we use a video diffusion method, which models temporal consistency between frames as well as producing high quality spatial detail.

To tackle these problems, we present CPMGEM-hourly, an emulator of sub-daily rainfall from km-scale simulations, based on a video diffusion model. Our emulator uses daily-mean atmospheric variables as inputs, which are readily attainable from climate models, and overcomes the previously mentioned difficulties.
Such an emulator makes rainfall predictions based on high-quality physical models, including future climate changes, sufficiently affordable to study rainfall extremes and to drive national-scale physically-based hydrological and flood models to study flood extremes.
In this work, we test our approach’s ability to recreate realistic spatial and temporal rainfall structures and the intensity distribution, as well as changes through the 21st century.

\section{Method}
\label{sec:method}

% Schematic figure?

\paragraph{Data} We train the model on the UK Climate Projections 2018 Local dataset (UKCP Local), which used the Met Office's physically-based UK convection-permitting model (CPM), which has 2.2km grid spacing \cite{kendonVariabilityConcealsEmerging2023, kendonUpdateUKCPLocal2021}.
This consists of 12 ensemble members covering three 20 year periods (1981-2000, 2021-2040 and 2061-2080). The simulations were driven using a GCM with 60km grid spacing.

We emulate hourly precipitation from the CPM over England and Wales.
As in work by \citet{addisonMachineLearningEmulation2026} to emulate daily-mean precipitation, we coarsen the precipitation to 8.8km resolution to capture the reasonably well-resolved scales of the CPM.

We use a ``perfect model'' framework, meaning we use coarsened CPM variables as inputs and CPM precipitation as our target.
The input variables are coarsened to match the 60km grid spacing of the driving GCM.
We use the same coarse, daily-mean atmospheric variables as inputs as \citet{addisonMachineLearningEmulation2026}, over the same spatial domain as the target precipitation: pressure at mean sea level; and specific humidity, relative vorticity and temperature at 850, 700, 500 and 250 hPa.

% The coarse inputs are available on Zenodo \cite{addison2025cpmgemdataset} and hourly precipitation is available from the CEDA Archive \cite{UKCPLocalCEDA}. An AI-ready dataset combining these for easier use will be released later.

\paragraph{Emulator}
Our emulator, ``CPMGEM-hourly'', is based on Latte \cite{ma2025latte}, a video generation model that uses a latent diffusion transformer. In this work the encoding to a latent space has been deactivated, as we found this resulted in better performance.
% Latte performed well on the VBench video generative model benchmark suite \cite{huang2024vbench}, and is open source. 
The model was adapted to produce 24 frames of output (one for each hour of the day), with output conditioned on the daily, coarse-scale input variables.
We train on 4 Nvidia H100 GPUs and perform inference with a single GPU.
% The transformer contains a mix of temporal and spatial transformer blocks and uses sinusoidal timestep embeddings.
The results in this work are based on the extra large (XL) configuration with a patch size of 8 from \citet{ma2025latte}, which has \(\sim\)675 million parameters.
For both training and sampling, the emulator uses 2000 diffusion steps.
% Exploration over choices of model size and numbers of diffusion steps found that reducing either produced modestly worse performance (not shown).
The emulator was trained for 150 epochs.%, which was found to optimise performance metrics.
 
%The coarse-scale inputs are interpolated using a nearest-neighbour approach to match the resolution of and then concatenated with the 64-by-64 resolution high-resolution target precipitation.

\paragraph{Data processing}
We used 70\% of the data (504 years) for training and evaluated the output on 15\% (108 years), using the same approach as \citet{addisonMachineLearningEmulation2026} to obtain a balanced mix of ensemble members, time periods and seasons within each dataset. The remaining 15\% is reserved for testing for final model development in future work. We generate one sample per day for evaluating results in \autoref{sec:res}.

Precipitation and input variables are transformed using the method of \citet{addisonMachineLearningEmulation2026}, except the fourth root of precipitation is used rather than the square root, as this was found to give better performance for hourly precipitation.
% This may be because the distribution of hourly precipitation is more heavily skewed than that for daily data.

\section{Results}
\label{sec:res}

\begin{figure}[htbp]
    \centering
    \includegraphics[width=0.9\linewidth]{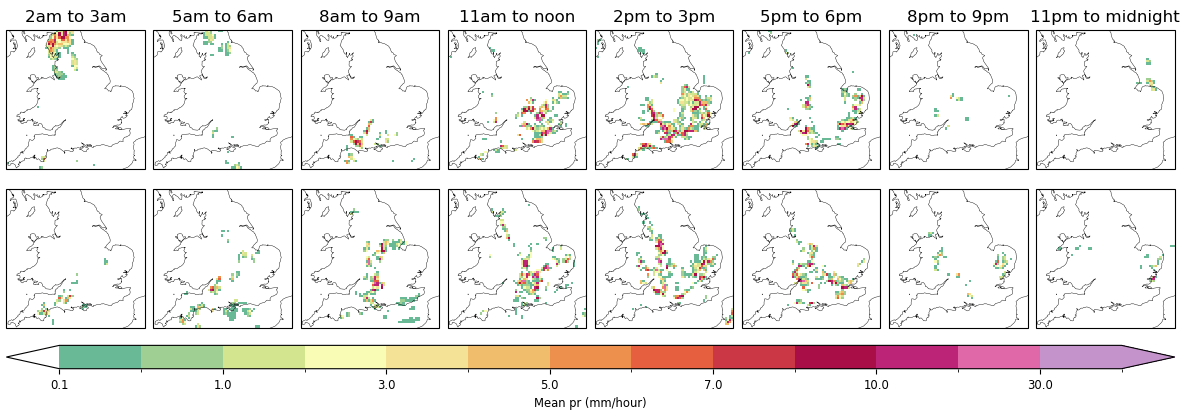}
    \caption{\textbf{Examples of predictions of hourly precipitation.} Example of a day with summer convective showers, showing hourly precipitation for eight hours through the day (2am to 3am, 5am to 6am, 8am to 9am, etc...). Top row shows output of the CPM target. Bottom row shows a sample generated by CPMGEM-hourly. Note that the CPM precipitation is effectively highly stochastic, and so samples from CPMGEM-hourly are not expected to match it in full detail, but to capture the overall statistics and have realistic spatial and temporal structure.}
    \label{fig:samples}
\end{figure}

Eight hours of example output are shown \autoref{fig:samples} for a summer day with convective showers. The top row shows the precipitation from the CPM. The bottom row shows a sample from CPMGEM-hourly conditioned on the corresponding coarse daily-mean inputs. An exact match between the two is not expected due to the large effectively-stochastic component in both the spatial and temporal downscaling \cite{addisonMachineLearningEmulation2026}, but the emulator is able to recreate the evolution of the pattern of small high-intensity convective events and the timing of their occurrence seen in the CPM in this example. 
% The emulator misses showers in northern England between midnight and 6am, but this may be due to these not being predictable from the daily-mean inputs without being given information on the initial rainfall. 
See \autoref{fig:app:rapsd} in Supplementary Information for radially averaged power spectral density distributions for the CPM and emulator samples for evidence that the emulator produces structures with similar variability at all spatial scales to the CPM across the full validation dataset.

% % % \begin{wrapfigure}{r}{0.7\linewidth}
% \begin{figure}[htbp]
%     \centering
%     \includegraphics[width=0.7\linewidth]{figures/freq.png}
%     \caption{\textbf{Histograms of precipitation values} in the CPM (grey) and CPMGEM-hourly (blue line). Note the vertical axis is logarithmic.}
%     \label{fig:freq}
% \end{figure}
% % % \end{wrapfigure}

% \begin{wrapfigure}{r}{0.7\linewidth}
\begin{figure}[htbp]
    % \centering
    \begin{subfigure}[t]{0.65\linewidth}
    % \centering
    \includegraphics[width=0.95\linewidth]{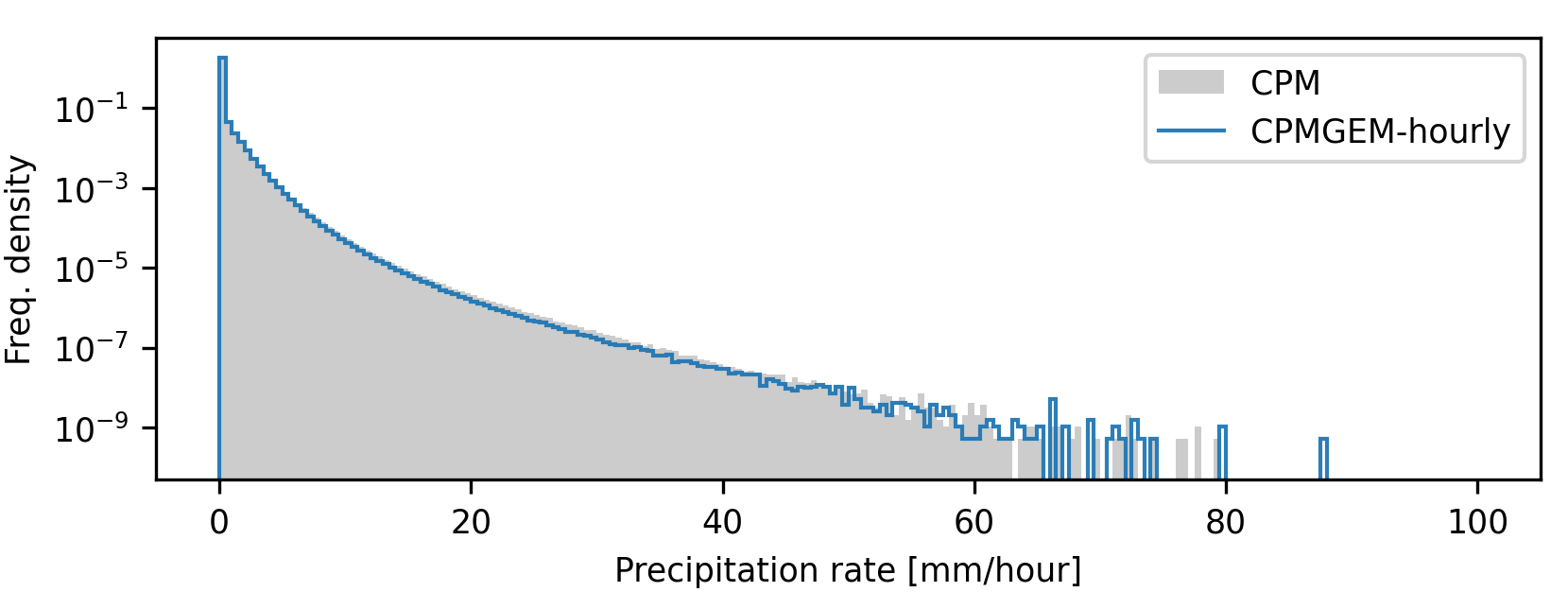}
    \caption{}
    \label{fig:freq}
    \end{subfigure}
    \begin{subfigure}[t]{0.4\linewidth}
    \includegraphics[width=\linewidth]{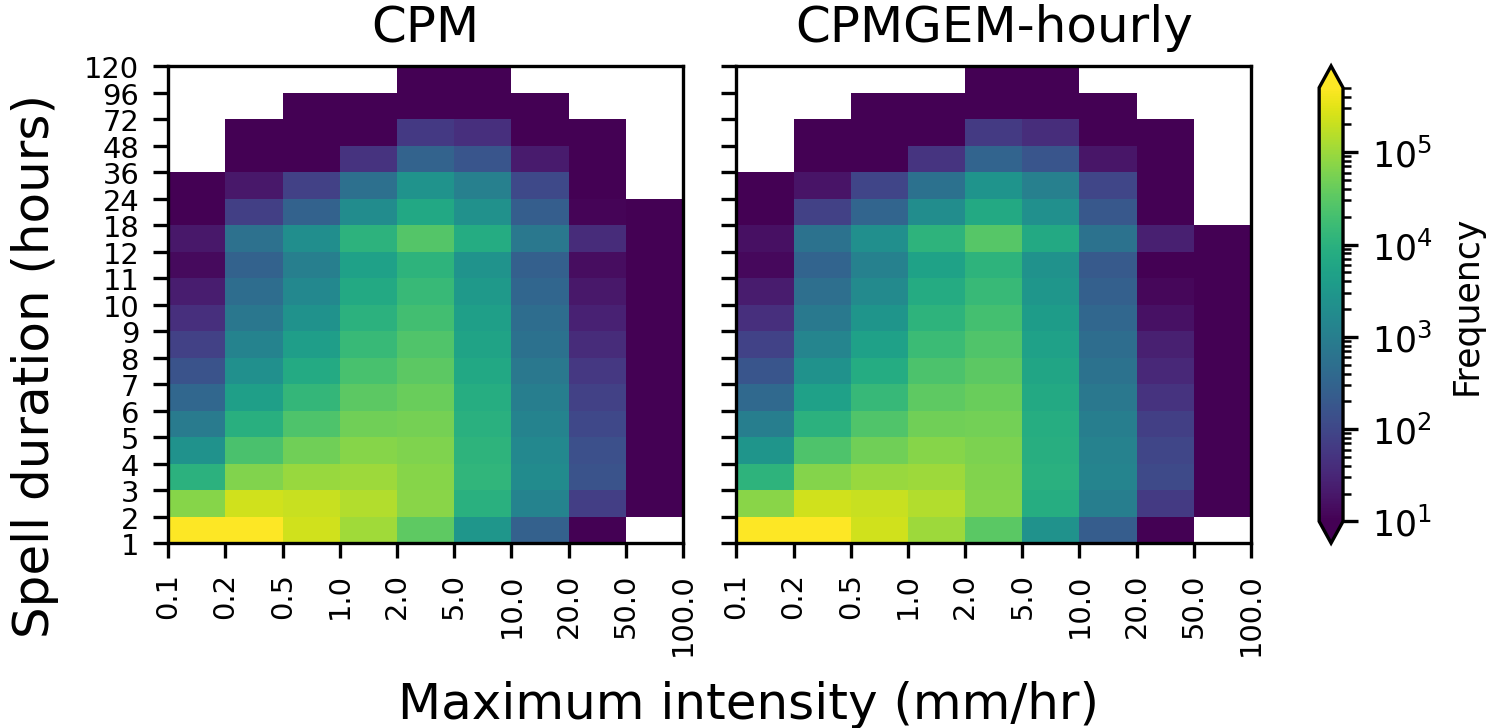}
    \caption{}
    \label{fig:idf}
    \end{subfigure}
    \caption{(a) \textbf{Histograms of precipitation values} in the CPM (grey) and CPMGEM-hourly (blue line). Note the vertical axis is logarithmic. (b) \textbf{2D frequency counts of wet spell duration and peak intensity} for the CPM (left) and CPMGEM-hourly (right).}
    \label{fig:freqandidf}
\end{figure}
% \end{wrapfigure}

\autoref{fig:freq} shows the frequency distribution of hourly precipitation across all grid boxes and hours for the CPM and CPMGEM-hourly.
The distribution of CPMGEM-hourly matches the CPM distribution closely, right up to the highest CPM intensities in the validation dataset (\(\sim\)80mm/hour). 
CPMGEM-hourly also captures the fact that only a small proportion of hours are ``wet'' (\(\geq\)0.1 mm/hour), closely matching the proportion of the CPM (10.5\% annually for both CPM and emulator), including the seasonal variation (from 15.6\% in winter to 5.8\% in summer, again for both). 
\autoref{fig:app:biases} shows relative biases in the mean, standard deviation and 99.9th percentile over the spatial domain. Their root mean squares over all grid boxes are 3.7\%, 5.3\% and 6.7\% respectively.
%Regarding the typical biases in precipitation statistics across the domain, the root mean square of the relative biases for the mean, standard deviation and 99.9th percentile are 3.7\%, 5.3\% and 6.7\% respectively (see \autoref{fig:app:biases} for spatial plots.).

% \begin{wrapfigure}{r}{0.5\linewidth}
% \begin{figure}[htbp]
%     \centering
%     \includegraphics[width=0.6\linewidth]{figures/idf.png}
%     \caption{\textbf{2D histograms of wet spell duration and peak intensity} for the CPM (left) and CPMGEM-hourly (right).}
%     \label{fig:idf}
% \end{figure}
% \end{wrapfigure}

% \begin{figure}[htbp]
%     \begin{subfigure}[t]{0.5\linewidth}
%     \centering
%     \includegraphics[width=\linewidth]{figures/idf.png}
%     \caption{}
%     \label{fig:idf}
%     \end{subfigure}
%     \begin{subfigure}[t]{0.55\linewidth}
%     \includegraphics[width=0.95\linewidth]{figures/ccs-seasonal-q999.png}
%     \caption{}
%     \label{fig:ccs}
%     \end{subfigure}
%     \caption{(a) \textbf{2D frequncy counts of wet spell duration and peak intensity} for the CPM (left) and CPMGEM-hourly (right). (b) \textbf{Changes in seasonal 99.9th percentile precipitation from 1981--2000 to 2061--2080.} Top row shows results for winter and bottom row shows results for summer. The first twos columns are the changes between 1981--2000 and 2061--2080 for CPM and CPMGEM-hourly, respectively. The third column is the difference between the two changes. Shown as percentages of the 1981-2000 CPM value.}
%     \label{fig:idfandccs}
% \end{figure}

It is also important to capture the joint distribution of intensity and duration of precipitation systems
%An important temporal characteristic of hourly precipitation is the relationship between the duration and intensity of wet spells at a grid box 
\cite{kendonRealismRainfallVery2012}.
For each grid box, wet spells are defined as continuous periods of precipitation above 0.1 mm/hour, and the duration and maximum intensity are calculated for each spell.
\autoref{fig:idf} shows the joint frequency counts of these variables.
These match closely between the CPM and CPMGEM-hourly, including the tendency for higher intensity spells to last longer.

\paragraph{Climate change signal} Representing future climate change signals is a key goal of physical model emulation. It is important for impacts modelling that changes in heavy precipitation are reproduced well. This has been a particularly difficult challenge for previous work on emulating regional models \cite{addisonMachineLearningEmulation2026, rampal2026intercomparisongenerativemachinelearning}.

% \begin{figure}[htbp]
\begin{wrapfigure}{r}{0.55\linewidth}
    \centering
    \includegraphics[width=\linewidth]{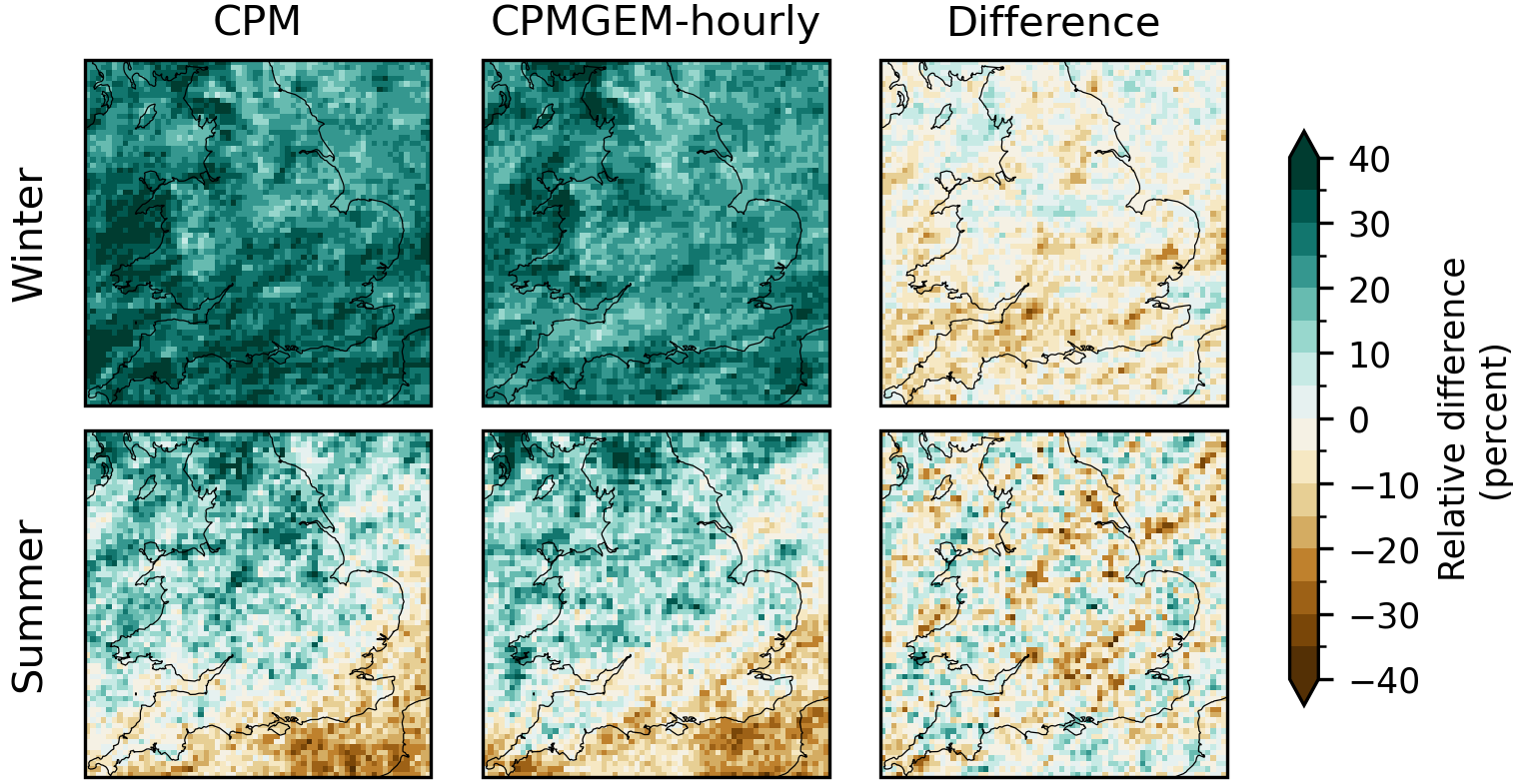}
    \caption{\textbf{Changes in seasonal 99.9th percentile precipitation from 1981--2000 to 2061--2080.} Top row shows results for winter and bottom row shows results for summer. The first twos columns are the changes between 1981--2000 and 2061--2080 for CPM and CPMGEM-hourly, respectively. The third column is the difference between the two changes. Shown as percentages of the 1981-2000 CPM value. \vspace{40pt}}
    \label{fig:ccs-q999}
% \end{figure}
\end{wrapfigure}

\autoref{fig:ccs-q999} shows that CPMGEM-hourly is able to reproduce the magnitude and spatial patterns of the relative change of 99.9th percentile of precipitation between the 1981--2000 and 2061--2080 time periods, including capturing differences between seasons. In winter (top), the emulator predicts an increase in the intensity at this percentile throughout the domain, very similarly to the CPM. In summer (bottom), the emulator also reproduces well the more complex spatial pattern of the change, with the intensity set to decrease in the south-eastern corner of the domain and increase elsewhere, with there being no large-scale spatial pattern in the difference in this change between the emulator and CPM. \autoref{fig:app:ccs-hist} demonstrates the emulator samples capture the change in the frequency distribution of precipitation between different time periods, including for very high intensity events.

\section{Conclusion}

We have demonstrated the skill of the CPMGEM-hourly emulator of precipitation output from a km-scale regional climate downscaling model. This indicates that video diffusion methods have substantial potential for use in climate downscaling. This advances upon previous work by generating sub-daily (hourly) precipitation with realistic spatial and temporal structure at high (8.8km) spatial resolution, conditioned on daily, coarse-resolution (60km) atmospheric variables, rather than only generating daily means as in \cite{addisonMachineLearningEmulation2026}.
The emulator reproduces important characteristics of the spatial and temporal structure of the CPM precipitation well, including the spatial spectral power distribution and the joint frequency distribution of wet spell durations and intensities.
The emulated samples recreate the precipitation intensity distribution, including high-intensity events, as well as changes in the intensity of heavy precipitation through the 21st century.

The emulator has much lower computational cost than the CPM (the configuration used for this work can generate $\sim$210 days of hourly samples per hour on a single GPU). %, though further work is needed to improve efficiency as we continue to
Further work will increase the spatial resolution and domain size of the output to cover the whole UK.
%This may be possible by reducing the size of the neural network and using the latent diffusion approach.% to reduce the computational cost and GPU memory requirements of the emulator.
Future planned applications include downscaling very large climate simulation ensembles of the 21st century to quantify changes in extreme rainfall statistics and to better inform resilience and climate adaptation planning.
Extreme events with frequencies of only once per $\sim$100--1000 years are highly important, and evaluation will be extended to quantify model performance on such cases. The emulator output will be further applied to drive national-scale hydrological and flood modelling systems, and evaluated on the realism of resulting simulated streamflow and flood events.
%The use of these rainfall predictions with hydrological and flood models will need to be tested, which will require emulator to also output how much water might come out of the ground in the form of potential evapotranspiration.

\clearpage
\bibliography{references}
\clearpage

% Figures become A.1, A.2, B.1
\renewcommand{\thefigure}{A.\arabic{figure}}
\setcounter{figure}{0}
\renewcommand{\appendixtocname}{Supplementary Information}

\appendix
\appendixpage

\begin{figure}[htbp]
    \centering
    \includegraphics[width=0.8\linewidth]{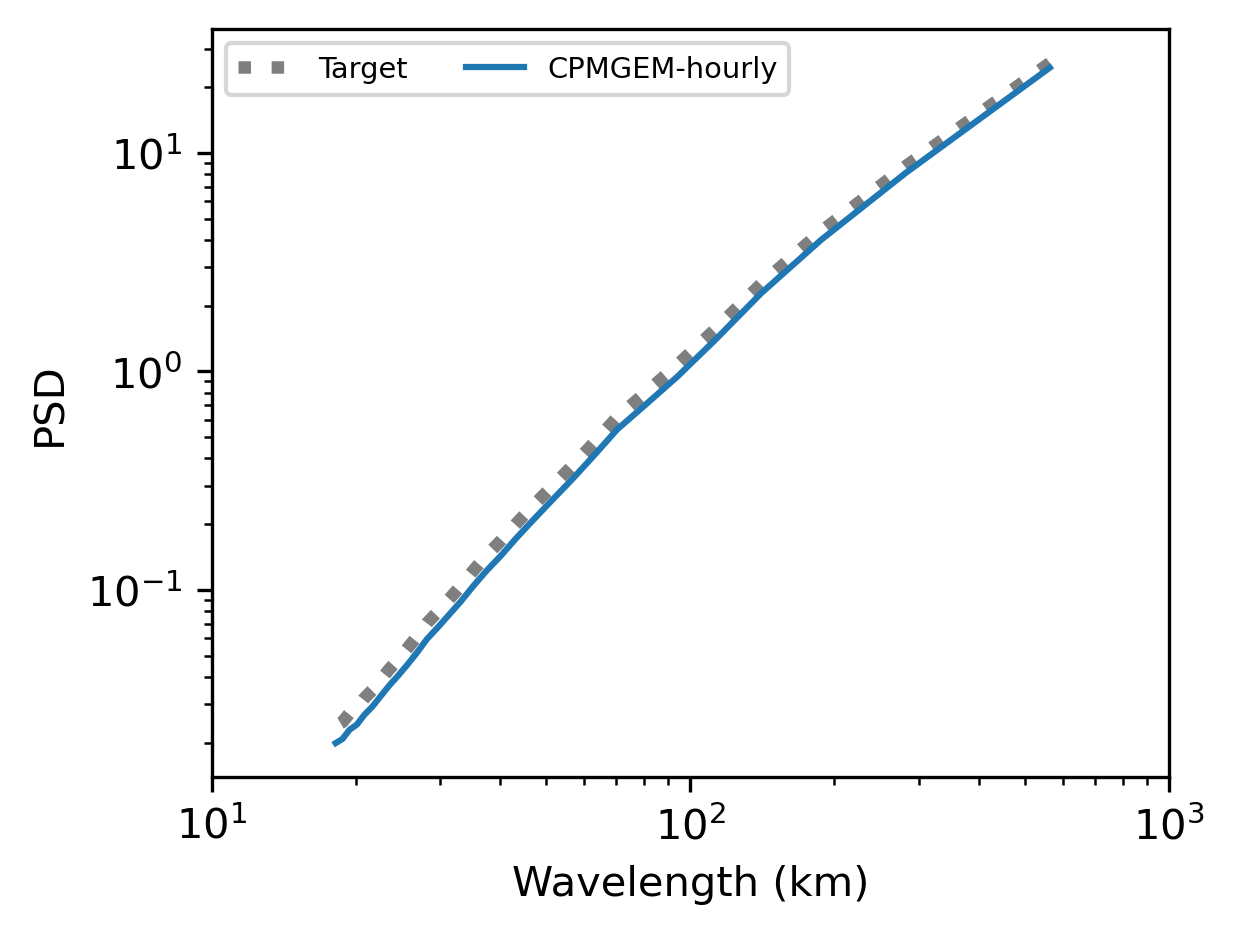}
    \caption{\textbf{Radially averaged spatial power spectral density (RAPSD)}. Shows the target CPM precipitation (grey dashes) and the emulator samples, CPMGEM-hourly (blue line).}
    \label{fig:app:rapsd}
\end{figure}

\begin{figure}[h!]
    \centering
    \includegraphics[width=0.8\linewidth]{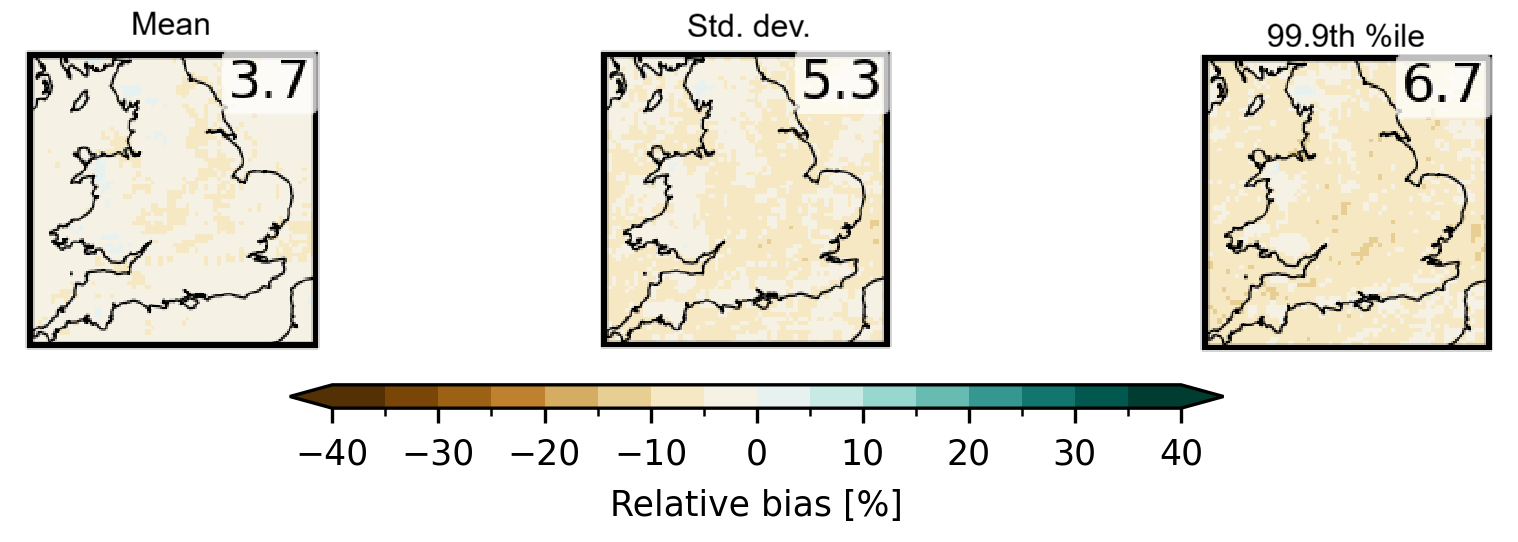}
    \caption{\textbf{Spatial plots of relative biases}. Relative bias of mean (left), standard deviation (middle) and 99.9th percentile (right), as a percentage of the CPM statistic. Root mean squared bias (\%) over the domain is shown in top right of each spatial plot.}
    \label{fig:app:biases}
\end{figure}

\begin{figure}[htbp]
    \centering
    \includegraphics[width=0.8\linewidth]{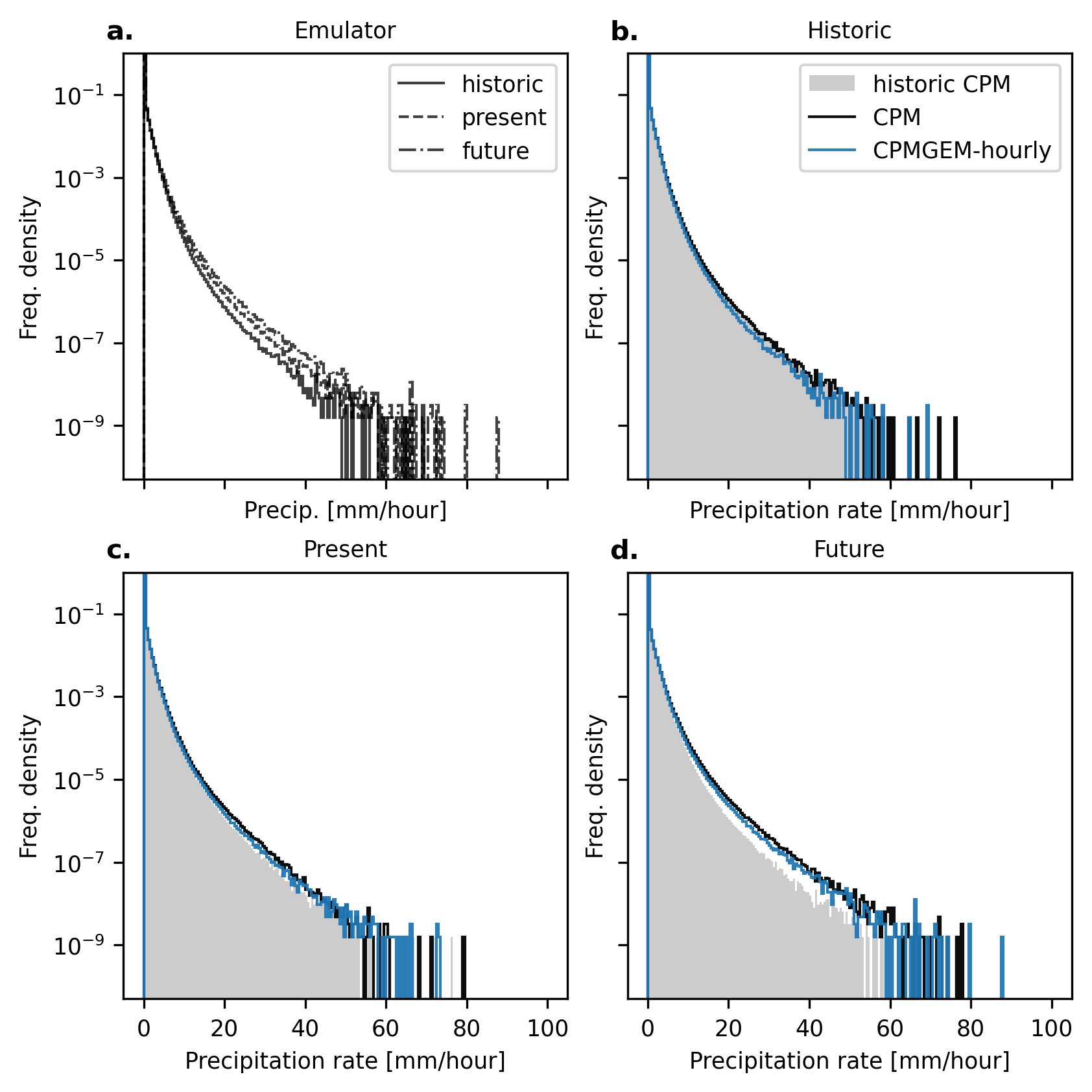}
    \caption{\textbf{Climate change effect on precipitation frequency distributions.} (a) shows the different CPMGEM-hourly frequency densities for the three different time periods: Historic (1981--2000, solid black), Present (2021--2040, dashed black) and Future (2061-2080, dotted black). (b) shows a comparison of the frequency density histogram of CPMGEM-hourly (blue) with the CPM (black) for the Historic time period. (c) and (d) show the same as (b) but for Present and Future time periods respectively. In (b), (c) and (d) the filled grey histogram shows the CPM precipitation frequency distribution from the Historic time period to highlight the change in the precipitation distribution between the time periods.}
    \label{fig:app:ccs-hist}
\end{figure}

\end{document}